\documentstyle[amsfonts]{amsart}

\def\bull{\vbox{\hrule\hbox{\vrule\kern3pt\vbox{\kern6pt}\kern3pt\vrule}\hrule}}

\def\a{\alpha}

\def\b{\beta}
\def\c{\gamma}

\def\O{\Omega}
\def\l{\lambda}

\def\top{\text{top}}

\def\RM{\Bbb R}
\def\CM{\Bbb C}

\def\gm{\bold g}

\def\<#1,#2>{\langle #1,#2\rangle}
\def\TR{\text{tr}\,}
\def\dep(#1,#2){\text{det}_{#1}#2}
\def\norm(#1,#2){\parallel #1\parallel_{#2}}
\def\Z{{\mathbb Z}}
\def\Cstar{U(1)}

\def\C{{\mathbb C}}
\def\U{{\mathcal U}_{res}}
\def\Uhat{\hat{\mathcal U}_{res}}
\def\Gr{{\mathcal G}r}
\def\d{{\text d}}
\def\WZW{\text{WZW}}
\def\hol{\text{hol}}

\numberwithin{equation}{section}

\theoremstyle{plain}
\newtheorem{theorem}{Theorem}[section]

\newtheorem{proposition}{Proposition}[section]

\theoremstyle{definition}
\newtheorem{definition}{Definition}[section]

\theoremstyle{remark}
\newtheorem{example}{Example}[section]

\begin{document}
\baselineskip=16pt

\title[BUNDLE GERBES APPLIED TO QUANTUM FIELD THEORY]{BUNDLE GERBES APPLIED
TO QUANTUM FIELD THEORY}

\author[A.L. Carey]{Alan L. Carey}
\address{Department of Mathematics\\ University of Adelaide\\ Adelaide SA
5005\\ Australia}
\email{acarey\@maths.adelaide.edu.au}

\author[J. Mickelsson]{Jouko Mickelsson}
\address{Department of Theoretical Physics\\ Royal Institute of
Technology\\
S-10044 Stockholm\\ Sweden}
\email{jouko\@theophys.kth.se}

\author[M.K. Murray]{Michael K. Murray}
\address{Department of Mathematics\\ University of Adelaide\\ Adelaide SA
5005\\ Australia}
\email{mmurray\@maths.adelaide.edu.au}

\begin{abstract}
This paper reviews recent work on a new geometric object called a
bundle gerbe and discusses some new
 examples arising in quantum field theory.
One application is to
an Atiyah-Patodi-Singer index theory construction
of the bundle of fermionic Fock spaces parametrized by vector potentials
in
odd space dimensions and  a proof that this leads in a simple manner to
the
known Schwinger terms (Mickelsson-Faddeev cocycle) for the gauge group
action.
This gives an explicit computation of the Dixmier-Douady
class of the associated bundle gerbe.
 The method works also in other cases of fermions
in external fields (external gravitational field, for example) provided
that the APS theorem can be applied; however, we have worked out the details
only in the case of vector potentials.
Another example, in which the bundle gerbe curvature
plays a role, arises from  the WZW model on
Riemann surfaces. A further example is  the `existence of string structures'
question. We conclude by
showing how global Hamiltonian anomalies fit within this framework.
\end{abstract}

\maketitle

\section{Introduction}

In \cite{Br} J.-L.
 Brylinski describes Giraud's theory of gerbes
and gives some applications particularly to geometric
quantisation. Loosely speaking
a gerbe over a manifold $M$ is a sheaf of groupoids over $M$. Gerbes,
via their Dixmier-Douady class,
provide a geometric realisation of the elements of $H^3(M, \Z)$ analogous
to the way that line bundles provide, via their Chern class,
a geometric realisation of the elements of $H^2(M, \Z)$.

There is a simpler way of achieving this
end which, somewhat surprisingly, is nicely adapted to
the kind of geometry arising in quantum field theory
applications. For want of a better name these
objects are called {\it bundle gerbes} and are introduced in
\cite{Mu}.
All this talk of sheaves and groupoids sounds
very abstract. In this article  we want to illustrate the importance and
usefulness of bundle gerbes by describing five natural examples
arising in different parts of quantum field theory. These are:

\begin{itemize}
	\item  string structures,

	\item  $\U$ bundles,

     \item  the Wess-Zumino-Witten action

	\item  local Hamiltonian anomalies (the Mickelsson-Faddeev cocycle).

\item global Hamiltonian anomalies

\end{itemize}

The value of the bundle gerbe picture can be seen from
the fourth and fifth examples: they provide a geometric meaning
to these anomalies which previously have been thought of
as associated with cocycles on the gauge group.

 Just as a gerbe is a sheaf of groupoids a bundle
gerbe is essentially a bundle of groupoids.  A bundle
gerbe has associated to it a three-class also called
the Dixmier-Douady class. Every bundle gerbe gives rise to
a gerbe with the same Dixmier-Douady class.
Bundle gerbes behave in many ways like line bundles.
There   is a notion of  a trivial bundle gerbe and a bundle gerbe
is trivial if and only if its Dixmier-Douady class vanishes.
One can form the dual and tensor products of bundle gerbes and
the Dixmier-Douady class changes sign on the dual and is additive
for tensor products.  Every bundle gerbe admits a bundle
gerbe connection which can be used to define a three form
on $M$ called the curvature of the connection and which
is a de Rham representative for $2\pi i$ times the non-torsion part of the
Dixmier-Douady class.  A difference with the line bundle
case is that one needs to choose not just the connection
but an intermediate two-form called the curving to define
the curvature.  There is a notion of holonomy
for a connection and curving but now it is associated to a
two-surface rather than a loop. We exploit this in our
description of the  Wess-Zumino-Witten action.
There is a local description of
bundle gerbes in terms of transition functions and a corresponding
\v Cech definition of the Dixmier-Douady class. Finally
bundle gerbes can be pulled back and there
 is a universal bundle gerbe and an associated
classifying theory.

The one significant difference between the two structures;
 lines bundles and bundle gerbes; is that
two line bundles are isomorphic if and only if their Chern
classes are equal, whereas two bundle gerbes which are
isomorphic have the same Dixmier-Douady class but the
converse is not necessarily true. For bundle
gerbes there is a weaker notion of isomorphism
called  {\em stable isomorphism} and
 two bundle gerbes are stably isomorphic if
and only if they have the same Dixmier-Douady class \cite{MuSt}.
The reader with some knowledge of groupoid or category theory
will recall that often the right concept of equal
for categories is that of equivalence which is
weaker than isomorphism.   A similar situation
arises for bundle gerbes essentially because they are bundles
of groupoids.

A common thread in the examples we consider is the relationship
between bundle gerbes and central extensions.  Because
group actions in quantum field theory are only
projectively defined one often needs to consider
the so-called `lifting problem' for principal
$G$ bundles where $G$ is   the quotient in a central
extension:
$$
U(1) \to \hat G \to G.
$$
The lifting problem starts with a   principal $G$ bundle and seeks
to find
a lift of this to a principal $\hat G$ bundle. The obstruction
to such a lift is well known to be a class in $H^3(M, \Z)$.
The connection with bundle gerbes arises because there is a
 so-called lifting bundle gerbe,
which   is trivial
if and only if the principal $G$ bundle lifts and its Dixmier-Douady
class is the three class obstructing the lifting. In the
first  and third examples $G$ is the loop group with its standard
central extension (the Kac-Moody group), in the second example
$G$ is the restricted unitary group $\U$ with its
canonical central extension,
and in  the fourth and fifth examples
it is a gauge group of a chiral gauge theory.

  It is important
to note, and the central point of this paper, that  the
bundle gerbe description arises naturally and usefully in these examples
and is not just a fancier way of describing the
lifting problem for principal bundles.

In summary form we present the basic theory of bundle
gerbes in Section \ref{bg}. This is followed by the examples:
the gerbes arising from
global Hamiltonian anomalies are described in Section 6,
the lifting problem for the restricted unitary bundles and
string structures is in Section \ref{Ures},
the Mickelsson-Faddeev cocycle (local
Hamiltonian anomalies) is in Section \ref{FM} and  the
geometric interpretation provided by bundle gerbes of the Wess-Zumino-
Witten term is in Section \ref{WZW}.

Section \ref{FM} is a summary of \cite{CaMiMu}
and also forms part of a previous short review \cite{CaMiMu1}.
We include it because it is essential for the understanding
of the later sections.
The material in Sections 3, 5 and 6 is new. Section 3 may be skipped on
first
reading (it is a bit technical). Section 5 depends a little on Section 4
and Section 6 on parts of both Sections 4 and 5.

We conclude this introduction by
remarking that there are `bundle n-gerbes' associated
with classes in $H^{n+2}(M,\Z)$. Examples are given in \cite{CaMuWa}
and the general theory in \cite{St} however it would take
us too far afield to describe them here.

\section{Bundle gerbes}
\label{bg}

This Section is a review: we describe the basic theory of
bundle gerbes.  We will not prove any of the results
but refer the reader to \cite{Mu, CaMiMu, MuSt, CaMu}
and the forthcoming thesis of Stevenson \cite{St}.

\subsection{The definition and basic operations}
\label{defs}
Consider a submersion
$$
\pi \colon Y \to M
$$
and define $Y_m =  \pi^{-1}(m)$ to be the
fibre of $Y$ over $m \in M$.
Recall that  the fibre product
$Y^{[2]}$ is a new submersion over $M$ whose
fibre at $m$ is $Y_m \times Y_m$.

A bundle gerbe $(P, Y)$ over $M$ is defined to be a choice of a submersion
$Y
\to M$ and a $\Cstar$ bundle $P \to Y^{[2]}$ with a product, that is,
a $\Cstar$ bundle isomorphism $P_{(y_1, y_2)} \otimes P_{(y_2, y_3)}
 \to P({y_1, y_3)}$. The product is required to be
associative whenever triple products are defined.

\begin{example}
Let $Q \to Y$ be a principal $\Cstar$ bundle.
Define $$P_{(x, y)} =  Aut_{\Cstar}(Q_x, Q_y) =  Q_x^* \otimes Q_y$$
Then this defines a  bundle gerbe  called the trivial bundle gerbe.
\end{example}

A morphism of bundle gerbes $(P, Y)$ over $M$ and $(Q , X) $ over $ N$ is
 a triple
of maps $(\a, \b, \c)$. The map $\beta \colon Y \to X$ is required to
be a morphism of the submersions $Y \to M$ and $X \to N$ covering
$\gamma \colon M \to N$. It therefore induces a morphism $\beta^{[2]}$

 of the submersions $Y^{[2]} \to M$ and $X^{[2]} \to N$. The  map
$\alpha$ is required to be a morphism of $\Cstar$ bundles covering
$\beta^{[2]}$  which commutes with the product.
A morphism of bundle gerbes over $M$ is a morphism of bundle
gerbes for which $M=N$ and $\gamma$ is the identity on $M$.

Various constructions are possible with bundle gerbes.
We can define a pull-back and product as follows.
If $(Q, X)$ is a bundle gerbe over $N$ and $f \colon M \to N$
is a map then we can pull back the submersion $X\to N$ to
obtain a submersion $f^*(X) \to M$ and a morphism of
submersions $f^{*} : f^{*}(X) \to X$ covering $f$.
This induces a morphism $(f^{*}(X))^{[2]} \to X^{[2]}$ and
hence we can use this to pull back the $\Cstar$ bundle
$Q$ to a  $\Cstar$ bundle $f^{*}(Q)$ say on $f^{*}(X)$.
It is easy to check that $(f^{*}(Q), f^{*}(X))$
is a bundle gerbe, the pull-back by $f$ of the gerbe
$(Q, X)$. If $(P, Y)$ and $(Q, X)$
are bundle gerbes over $M$ then we can form the fibre
product $Y\times_M X  \to M$ and then form a $\Cstar$ bundle
$P \otimes Q$ over $(Y\times_M X )^{[2]}$  which we call the
product of the bundle gerbes $(P, Y)$ and $(Q, X)$.

Notice that for any $m \in M$ we can define a {\em groupoid}
as follows. The objects of the groupoid are the
points in $Y_m = \pi^{-1}(m)$. The morphisms
between two objects $x, y \in P_m$ are the elements
of $P_{(x, y)}$. The bundle gerbe product defines
the groupoid product. The existence of identity and inverse
morphisms is shown in \cite{Mu}. Hence we can
think of the bundle gerbe as a family of groupoids,
parametrised by $M$.

\subsection{The Dixmier-Douady class and stable isomorphism}
\label{dd}
Let $\{U_\a\}$ be an open cover of $M$ such that
over each $U_\a$ we can find sections $s_\a \colon U_\a \to Y$.
Then over intersections $U_\a \cap U_\b$ we can define a map
$$
(s_\a, s_\b) \colon U_\a \cap U_\b \to Y^{[2]}
$$
which sends $m$ to $(s_\a(m), s_\b(m))$.  If we choose a
sufficiently nice cover we can then find maps
$$
\sigma_{\a\b} \colon U_\a \cap U_\b \to P
$$
such that $\sigma_{\a\b}(m) \in P_{(s_\a(m), s_\b(m))}$.
The $\sigma_{\a\b}$ are sections of the pull-back of $P$
by $(s_\a, s_\b)$.  By using the bundle gerbe multiplication
(written here as juxtoposition) we have that
$$
\sigma_{\a\b}(m) \sigma_{\b\c}(m)  \in P_{(s_\a(m), s_\c(m))}
$$
and hence can be compared to $\sigma_{\a\c}(m)$. The
difference is an element of $\Cstar$ defined by
$$
\sigma_{\a\b}(m) \sigma_{\b\c}(m) =  \sigma_{\a\c}(m) g_{\a\b\c}(m)
$$
and this defines a map
$$
g_{\a\b\c} \colon U_\a \cap U_\b \cap U_\c \to \Cstar.
$$
It is straightforward to check that this
is a cocycle and defines an element of $H^2(M, \underline{\Cstar})$
independent of all the choices we have made.  Here, if $G$ is a Lie
group we use the notation $\underline{G}$ for the sheaf of
smooth maps into $G$. It is a standard result
that the coboundary map
\begin{equation}
\label{coboundary}
\delta \colon H^2(M, \underline{\Cstar}) \to H^3(M, \Z)
\end{equation}
induced by the short exact sequence of sheaves
$$
0 \to \Z \to \underline{\CM} \to \underline{\Cstar} \to 0
$$
is an isomorphism. Either the class defined by $g_{\a\b\c}$ or its image
under
the coboundary map is called the
Dixmier-Douady class of the bundle gerbe. We denote
it by $\d(Q, Y)$.

The first important fact about the Dixmier-Douady class is
\begin{proposition}[\cite{Mu}]
A bundle gerbe is trivial if and only its Dixmier-Douady class
is zero.
\end{proposition}

Let $(P, Y)$ and $(Q, X)$ be bundle gerbes over $M$ and
$Z \to M$ be a map admitting local sections with
$f\colon Z \to Y$   a map commuting with projections to $M$.  Then
from \cite{Mu} we have
\begin{theorem}[\cite{Mu}]
\label{th:dd}
If $P$ and $Q$ are bundle gerbes over $M$ then
\begin{enumerate}
\item $d(P^*, Y) =  -d(P, Y)$
\item$d(P\otimes Q, Y\times_M X ) =  d(P, Y)
+ d(  Q, X)$, and
\item$d(f^*(P) , X) =  f^*(d(P, Y)) $
\end{enumerate}
\end{theorem}

Because pulling back the submersion does not
change the Dixmier-Douady class of a bundle gerbe
it is clear there are many non-isomorphic bundle
gerbes with the same  Dixmier-Douady class.
We define
\begin{definition}[\cite{MuSt}]
\label{stableiso}
Two bundle gerbes $(P, Y)$ and $(Q, Z)$ are
defined to be {\em stably isomorphic} if there are trivial bundle
gerbes $T_1$ and $T_2$ such that
$$
P \otimes T_1 =  Q \otimes T_2.
$$
\end{definition}
We have the following theorem:
\begin{theorem}[\cite{MuSt}]
For bundle gerbes $(P, Y)$ and $(Q, Z)$
the following are equivalent.
\begin{enumerate}
\item $P$ and $Q$ are stably isomorphic
\item $P \otimes Q^*$ is trivial
\item $d(P) =  d(Q)$.
\end{enumerate}
\end{theorem}

\subsection{Local bundle gerbes}
\label{local}
The notion of stable isomorphism is useful in understanding
the role of open covers in bundle gerbes.  Let $(P, Y)$
be a bundle gerbe and assume we have an open cover and various
maps $s_\a$, $\sigma_{\a\b}$ and $g_{\a\b\c}$ as defined in
subsection \ref{dd}.  Let $X$ be the disjoint union of all the
open sets $U_\a$. This can be thought of as all pairs $(\a, m)$
where $m \in U_\a$. There is a projection $X \to M$
defined by $(\a, m) \mapsto m$ which admits local sections.
 Moreover there is a map $s \colon X \to Y$
preserving projections defined by $s(\a, m) =  s_\a(m)$.

The pullback by $s$ of the bundle gerbe $(P, Y)$ is
stably isomorphic to $(P, Y)$ \cite{MuSt}. This pull-back consists
of a collection of $\Cstar$  bundles $Q_{\a\b} \to U_\a \cap U_\b$.
On triple overlaps there is a bundle map
$$
Q_{\a\b} \otimes Q_{\b\c} \to Q_{\a\c}
$$
which on quadruple overlaps is associative in the
appropriate sense. A completely local description of
bundle gerbes can be given in terms of open covers,
$\Cstar$ bundles on double overlaps and products on
triple overlaps \cite{St}. The results on stable
isomorphism tell us that this is equivalent to working
with bundle gerbes.

\subsection{Bundle gerbe connections, curving and curvature}
\label{conn}
Because $P \to Y^{[2]}$ is a $\Cstar$ bundle
it has connections. It is shown in \cite{Mu} that it
admits {\em bundle gerbe connections}
that is connections commuting  with the bundle gerbe product.
It is also shown there that the curvature $F$ of such a connection
must satisfy the `descent equation':
$$
F =  \pi_1^*(f) - \pi^*_2(f)
$$
for a two-form $f$ on $Y$. The two-form $f$ is not unique and
a choice of an $f$ is called a {\em curving} for the bundle gerbe
connection. It is then
easy to show that $df =  \pi^*(\omega)$ for some
three-form $\omega$ on $M$. In \cite{Mu} it is shown that
$\omega/{2\pi i}$ is a de Rham representative for the image of
the Dixmier-Douady class in real cohomology.

\subsection{The lifting bundle gerbe}
\label{lift}
Finally let us conclude this section
with the motivating  example of the
so-called lifting  bundle gerbe. That  is the
bundle gerbe arising from the lifting problem
for principal bundles. Consider a central extension of
groups:
\begin{equation}
1 \to \Cstar {\buildrel\iota \over \to } \  \hat G
{\buildrel p\over \to }\  G \to 1
\label{central}
\end{equation}
and a principal $G$ bundle $Y \to M$. Then it may happen that there is
a principal $\hat G$ bundle $\hat Y$ and a bundle map
$\hat Y \to Y$ commuting with the homomorphism $\hat G \to G$.
In such a case  $Y$ is said to lift to a $\hat G$ bundle.
One way of answering the question of when $Y$ lifts to a $\hat G$ bundle
is  to present $Y$ with transition functions $g_{\a\b}$ relative to a
cover $\{ U_\a\}$ of $M$. If the cover is sufficiently nice we can
lift the $g_{\a\b}$ to maps $\hat g_{\a\b}$ taking values in
$\hat G$ and such that $p(\hat g_{\a\b}) =  g_{\a\b}$.
These are candidate transition functions for a lifted bundle $\hat Y$.
However they may not satisfy the cocycle condition
$\hat{g}_{\a\b} \hat{g}_{\b\c} \hat{g}_{\c\a} =  1$
and indeed there is a $\Cstar$ valued map
$$
e_{\a\b\c} \colon U_\a \cap U_\b \cap U_\c \to \Cstar
$$
defined by $\iota(e_{\a\b\c}) = \hat g_{\a\b} \hat g_{\b\c} \hat g_{\c\a}$.
Because \eqref{central}  is a central extension it follows  that  $e_{\a
\b\c}$
is a cocycle and hence defines a class in $H^2(M, \underline{\Cstar})$.
Using the isomorphism \eqref{coboundary}
we define a class in $H^3(M, \Z)$ which is the
obstruction to solving the lifting problem for $P$.

Consider the fibre product of $P$ with itself.
There is a map $g\colon P^{[2]} \to G$ defined by
$p_1 =  p_2 g(p_1, p_2)$ and we can use this to pull back the
$U(1)$ bundle $\hat G \to G$ to define a $U(1)$ bundle
$Q \to P^{[2]}$. More concretely we have
$$
Q_{(p_1, p_2)} =  \{ g \in \hat G \mid p_1 p(g) =  p_2 \}.
$$
The group product on $\hat G$ induces a bundle gerbe product on $Q$.
It is shown in \cite{Mu} that the bundle gerbe $Q$ is trivial
  if and only if the bundle $P$ lifts to $\hat G$ and moreover the
Dixmier-Douady class of $(Q, P)$ is the class defined in the
preceeding paragraph.

\section{The Wess-Zumino-Witten term}
\label{WZW}
In  quantum field theory the path integral can have
contributions that are topological in nature. Often
these arise as the {\em holonomy} of a connection.
For example if $L \to M$ is a hermitian line bundle
we can consider the Hilbert space of all
$L^2$ sections of $L$ as a space of states. A
connection $\nabla$ on $L$ defines an operator
on states  by
$$
K_\nabla(\psi)(x)  =  \int_{M}
 \int_{\gamma \in W_{x, y}} P_\gamma(\nabla)(\psi(y)) dM
$$
where $W_{x, y}$ is the set of all paths from $x$ to $y$
and
$$
P_\gamma(\nabla) \colon L_y \to L_x
$$
is the operation of parallel transport along the curve $\gamma$
from the fibre of $L$ over $y$ to the fibre of $L$ over
$\gamma$. If $x= y$ then $P_\gamma(\nabla) \colon L_y \to L_x$
is an element of $\Cstar$ called $\hol(\gamma, \nabla)$,
 the {\em holonomy} of the connection $\nabla$ around the
loop $\gamma$. Assuming that $M$ is  simply connected every loop $\gamma$
bounds a disk $D$ and we have the fact that
\begin{equation}
\label{twohol}
\hol(\gamma, \nabla) =  \exp(\int_D F_\nabla)
\end{equation}
where $F_\nabla$,  is the curvature two-form
of $\nabla$.

The Wess-Zumino-Witten term is defined as
follows.  The space of states is replaced by the space of all maps (classical
field configurations) $\psi$  from a closed
Riemann surface $\Sigma$ into a compact Lie group $G$. Let $X$ be a
three-manifold whose boundary is $\Sigma$ and $\psi \colon \Sigma \to
G$. Then $\psi$ can be extended to a map $\hat \psi \colon X \to G$.
Let $\omega$ be a closed three-form on $G$ such that $\omega/2\pi i $
generates the integral cohomology of $G$.
The Wess-Zumino-Witten action of $f$ is then defined by
$$
\WZW(f) =  \exp{\int_X \hat \psi^*(\omega)}.
$$
It follows from the integrality of $\omega/2\pi i$ that this
is independent of the choice of extension $\hat \psi$.
Clearly
this is analogous to defining the holonomy by using the
right hand side of equation \eqref{twohol} as if one knew
nothing of connections, only that $F$ was a two-form such
that $F/2\pi i $ was integral.

It is natural to look for
the analogous left hand side of equation \eqref{twohol}
in the case of the Wess-Zumino-Witten term.
 In \cite{CaMu2}  an
interpretation of the Wess-Zumino-Witten term in terms of
holonomy of a connection on a line bundle over the loop group
of $G$ is given.  This essentially only worked for simple Riemann surfaces
such as spheres or cylinders. In
\cite{Ga} Gawedski gave a  construction that works for any Riemann surface.
Gawedski starts by
showing that isomorphism classes of line bundles with connection
are classified by certain Deligne cohomology groups which
can be realised in terms of \v Cech cohomology of an open cover
of $M$. It is then shown that if $M$ is a loop then this cohomology
group is $U(1)$ and the identification of isomorphism classes
with elements of $U(1)$ is just the  holonomy of the connection
around the loop.
This is then generalised to the Wess-Zumino-Witten case. For such
cohomology classes Gawedski shows that
it is possible to define a holonomy associated
to a closed Riemann surface in $M$.

Bundle gerbes give a geometric interpretation of Gawedski's
results. The Deligne cohomology class in question defines a stable
isomorphism class of a
bundle gerbe with connection and curving over $M$ and the element of
$\Cstar$ is the holonomy of the connection and curving over the
surface $\Sigma$.    In the case that $Y \to M$ admits local
sections a definition of the holonomy in terms of lifting $\Sigma$
to $Y$ is given in \cite{Mu}. In the present work we are interested
in more general $Y$, in particular a $Y$ arising from an open
cover,  so we give  an alternative definition of holonomy
of a bundle gerbe connection and curving.

To define the holonomy we need  the notion of a
stable isomorphism class of a bundle gerbe with connection and
curving.
To define this  let  $P \to Y$ be a line
bundle with connection $\nabla$ and curvature $F$.
The trivial gerbe $\delta(P)  \to Y^{[2]}$
 has a natural  bundle gerbe connection $\pi_1^*(\nabla) =
 \pi_2^*(\nabla)$ and curving $\pi_1^*(F) =
 \pi_2^*(F)$.  To extend  the definition
of stable isomorphism (definition \ref{stableiso}) to cover the case of
bundle gerbes with connection and curving we  assume that the trivial bundle
gerbes $T_1$ and $T_2$, in the definition,  have  connections arising in
this manner and that the
isomorphism  in definition \ref{stableiso} preserves connections. Then stable
isomorphism classes of bundle gerbes with connection and curving
are classified by the two dimensional Deligne cohomology
or the hyper-cohomology of the log complex of sheaves
\begin{equation*}
0 \to \Omega^\times \stackrel{d\log }{\to} \Omega^1 \to \Omega^2 \to 0 .
\end{equation*}
The hypercohomology of this sequence of sheaves can be
calculated by taking a Leray cover ${\mathcal U} = \{ U_\a \}$ of $M$
and
considering the
double complex:
\begin{equation}
\label{double_complex}
\begin{array}{ ccccccccc}
 \delta\uparrow &  & \delta\uparrow & & \delta\uparrow & & \delta\uparrow &
\\
C^2({\mathcal U}, \Omega^\times) & \stackrel{d\log }{\to} &
C^2({\mathcal U}, \Omega^1) & \to & C^2({\mathcal U}, \Omega^2) &
\stackrel{d }{\to} &
C^2({\mathcal U}, \Omega^3) & \stackrel{d }{\to} \\
 \delta\uparrow &  & \delta\uparrow & & \delta\uparrow & & \delta\uparrow
 &
\\
C^1({\mathcal U}, \Omega^\times) & \stackrel{d\log }{\to} &
C^1({\mathcal U}, \Omega^1) & \to & C^1({\mathcal U}, \Omega^2) &
\stackrel{d }{\to}  & C^1({\mathcal U}, \Omega^3) & \stackrel{d }{\to} \\
 \delta\uparrow &  & \delta\uparrow & & \delta\uparrow & & \delta\uparrow
 &  \\
 C^0({\mathcal U}, \Omega^\times) & \stackrel{d\log }{\to} &
C^0({\mathcal U}, \Omega^1) & \to & C^0({\mathcal U}, \Omega^2) &
\stackrel{d}{\to} &
C^0({\mathcal U}, \Omega^3) & \stackrel{d }{\to} \\
\end{array}
\end{equation}
The Deligne cohomology is the cohomology of the double complex
(\ref{double_complex}).
This
can be calculated by either of two spectral sequences which begin by
taking the cohomology along the rows or columns respectively.
If we assume that $M$ is a surface  $\Sigma$  and take the
cohomology of the columns we obtain the $E^2$ term of the
spectral sequence:
$$
\begin{array}{ cccc}
 0  &  0 &  0 &  0   \\
H^1(\Sigma, \Omega^\times) &  0  &0& 0\\
H^0(\Sigma, \Omega^\times) & H^1(\Sigma, \CM) & H^2(\Sigma, \CM) &
H^3(\Sigma, \CM)
\end{array}
$$
where we use the fact that the sheaves $\Omega^i$ have no cohomology and
that $$H^2(\Sigma, \Omega^\times) =  H^3(\Sigma, \Z) =  0$$ because
$\Omega^\times$ is the sheaf $\Cstar$.

The third cohomology is therefore the quotient of the
image of the inclusion
$$
\Z =  H^2(\Sigma, \Z) =  H^1(\Sigma, \Omega^\times) \to H^2(\Sigma,
\underline{\CM}) =  \CM
$$
induced by the second differential. It is straightforward
to check that this map is just the natural inclusion $\Z \to \CM$ and
hence the quotient is just $\Cstar$. We call the resulting
non-vanishing number attached to each connection and curving
the holonomy, $\hol(\Sigma, \nabla, f)$, where $\nabla$ is
the bundle gerbe connection and $f$ is the curving.
If $\psi \colon \Sigma \to M$ is a map then we can pull-back the
bundle gerbe, connection and curving to $\Sigma$  and we define
$\hol(\psi, \nabla, f)$ to be the holonomy of
the pulled-back connection and curving over $\Sigma$.

To calculate the holonomy we need to explain how to unravel
these definitions. Let us begin with a bundle gerbe $(Q, Y)$
and choose a Leray cover ${\mathcal U} =  \{ U_\a \}$ with sections
$s_\a \colon U_\a \to Y$. Choose sections $\sigma_{\a\b} \colon
U_\a \cap U_\b \to  Q$ as we did in subsection \ref{dd}
 and define $g_{\a\b\c}$ by
$$
\sigma_{\a\b}\sigma_{\b\c}^{-1} \sigma_{\c\a} =  g_{\a\b\c}.
$$
Let $A_{\a\b} $ be the pullback of the connection
one form on $Q$ by  $\sigma_{\a\b}$ and $f_\a$ the pullback of the
curving by $s_\a$. These satisfy
\begin{align*}
d\log{g_{\a\b\c}} &=  A_{\a\b} - A_{\b\c} + A_{\c\a} \\
d A_{\a\b} &=  f_\a - f_\b
\end{align*}
and hence $(g_{\a\b\c}, A_{\a\b}, f_\a )$ is an element
of the total cohomology of the complex \eqref{double_complex}.
Because $\Sigma$ is two dimensional the cocycle $g_{\a\b\c}$ is trivial
and  we can solve the equation
$$
g_{\a\b\c} =  h_{\b\c}h_{\a\c}^{-1} h_{\a\b}
$$
where $h_{\a\b} \colon U_\a\cap U_\b \to \Cstar$ and
hence
$$
d\log h_{\a\b} =  A_{\a\b} + k_\a - k_\b
$$
where the $k_\a$ are one-forms on $U_\a$. Hence
the two-form $f_\a - dk_\a$ on $U_\a$ agrees with the
two-form   $f_\b - dk_\b$ on  $U_\b$ on the overlap $U_\a \cap U_\b$
and hence defines a two-form on $\Sigma$. The exponential
of $2\pi$ times this two-form over $\Sigma$ is the
holonomy of the connection and curving.

It is straightforward to check that if we can extend the
map of $\Sigma$ into $M$ to a map of a manifold $X$ whose boundary
is $\Sigma$ then we obtain the analogue of \eqref{twohol}
$$
\hol(\nabla, f, \psi) =  \exp(\int_X \psi^*(\omega))
$$
where $\omega $ is the curvature of the connection
and curving (a 3-form).

In the case that $\Sigma$ has boundary one expects
a result analogous to parallel transport along a
curve $\gamma$. Gawedski shows that there is a naturally
defined line bundle $L$ over the space $LM$ of loops in $M$. The boundary
components $b_1, \dots, b_r$ of  $\Sigma$ define points in $LM$ and
Gawedski shows that the holonomy can be interpretated as
an element of
$$
L_{b_1} \otimes L_{b_2} \otimes \dots \otimes L_{b_r}
$$
for more details see \cite{Ga}.

\section{The Mickelsson-Faddeev cocycle}
\label{FM}
Let $M$ be a smooth compact connected
manifold without bound\-ary equi\-pped with a
spin structure. We assume that the dimension of $M$ is odd and equal to
$2n+1.$ Let $S$ be the spin bundle over $M,$ with fiber isomorphic
to  ${\CM}^{2^n}$. Let $H$ be the space of square integrable sections
of the complex vector bundle $S\otimes V,$ where $V$ is a trivial vector
bundle over $M$ with fiber to be denoted by the same symbol $V.$ The meas
ure
is defined by a fixed metric on $M$ and $V.$ We assume
that a unitary representation $\rho$ of a compact group $G$ is given in the
fiber. The set of smooth vector potentials on $M$ with values in the Lie
algebra $\gm$ of $G$ is denoted by $\mathcal A$.
The topology on $\mathcal A$ arises from an infinite family of Sobolev norms
which via the Sobolev embedding theorem give a metric equivalent to
that arising from the topology of uniform convergence of derivatives
of all orders.

For each $A\in \mathcal A$ there is a massless hermitean Dirac operator
 $D_A.$
Fix a
potential $A_0$ such that $D_A$ does not have zero as an eigenvalue
and let $H_+$ be the closed
subspace spanned by eigenvectors belonging to positive
eigenvalues of $D_{A_0}$ and $H_-$ its orthogonal complement
(with corresponding spectral projections $P_\pm$).
More generally for any potential $A$ and any real $\l$ not belonging
to the spectrum of
$D_A$ we define the spectral decomposition $H= H_+(A,\l)\oplus H_-(A,\l)$
with respect to the operator $D_A - \l.$ Let ${\mathcal A}_0$ denote the
set of
all pairs $(A,\l)$ as above and let
$$U_{\l}= \{A\in{\mathcal A}| (A,\l)\in{\mathcal A}_0\}.$$

Over the set $U_{\l\l'}= U_{\l}\cap U_{\l'}$ there is a canonical complex
line bundle, which we denote by $DET_{\l\l'}.$ Its fiber at $A\in U_{\l\l'}$
is the top exterior power
$$
DET_{\l\l'}(A) =  \wedge^{\top} (H_+(A,\l)\cap H_-(A,\l'))
$$
where we have assumed $\l<\l'.$ For completeness we put $DET_{\l\l'}=
DET_{\l'\l}^{-1}.$ Since $M$ is compact, the spectral subspace corresponding
to the interval $[\l,\l']$ in the spectrum is finite-dimensional and the
complex line above is well-defined.

It is known [Mi1, CaMu1] that there exists a complex line bundle $DET_{\l}$
over each of the sets $U_{\l}$ such that
	\begin{equation}
		DET_{\l'}=  DET_{\l}\otimes DET_{\l\l'}
		\label{det_eq}
	\end{equation}
over the set $U_{\l\l'}.$  In [CaMu, CaMu1] the structure of these line bundles
was studied with the help of bundle gerbes. In particular, there is an
obstruction for passing to the quotient by the group ${\mathcal G}$ of gauge
transformations which is  given by the Dixmier-Douady class of the
bundle gerbe.
(In [Mi1] the structure of the bundles and their relation to anomalies
was found by using certain embeddings to infinite-dimensional Grassmannians.)

We shall describe the computation in \cite{CaMiMu}
of the curvature of the (odd dimensional)
determinant bundles from Atiyah-Patodi-Singer index theory and how to obtain
the Schwinger terms in the Fock bundle directly from the local part of
the index density.

We may consider ${\mathcal A}_0$ as part
of a bundle gerbe over $\mathcal A$. The obvious map ${\mathcal A}_0 \to
{\mathcal A}$
 is a submersion.
For any $\lambda \in \RM$ we have a section $s_{\lambda} \colon
U_{\lambda } \to {\mathcal A}_0$ defined by $s_{\lambda}(A)  = (A, \lambda)$.
So we can apply the discussion in Section \ref{local}to obtain  the
disjoint union
 $$
 Y =  \coprod U_\lambda \subset \mathcal A \times \RM
 $$
 as the set of all $(A, \lambda) $ such that
$A \in U_\lambda$.  We topologize $Y$ by giving
$\RM$ the discrete topology. Notice that as a set
$Y$ is just ${\mathcal A}_0$ but the topology is different.
The identity map $Y \to {\mathcal A}_0$ is continuous.
It follows from the results of Section \ref{local} that  using either
topology  on ${\mathcal A}_0$ gives rise to stably isomorphic bundle
gerbes so we can work in either picture.
 An advantage of the open
cover picture is that the map $\delta$ introduced in
\cite{Mu} is then just the coboundary map in the
C\'ech de-Rham double complex. In the next section
${\mathcal A}_0$ can be interpreted in either sense.

If we restrict $\l$ to be rational then the sets $U_{\l}$
form a denumerable cover.
It follows  that the intersections  $U_{\l\l'}= U_{\l}\cap U_{\l'}$
also form a denumerable open cover. Similarly, we have an open cover by sets
$V_{\l\l'}=\pi(U_{\l\l'})$ on the quotient
 ${\mathcal X}={\mathcal A}/{\mathcal G}_e,$
where ${\mathcal G}_e$
 is the group of \it based \rm gauge transformations $g$, $g(p)=e$ the
identity at some fixed base point $p\in M.$  Here $\pi:\mathcal A\to X$ is the
canonical projection.

We now describe the bundle gerbe $J$ over $\mathcal A$ defined in [CaMu]
and extracted from the work of \cite{Se}.  First recall that there is an
equivalence between $U(1)$ bundles and hermitian line bundles, that
is complex line bundles with hermitian inner product on each fibre.
In one direction the equivalence associates to any hermitian
line bundle the $U(1)$ bundle of all vectors of unit norm.
It is possible to cast the definition of bundle gerbes in
terms of hermitian line bundles and indeed this was done in \cite{CaMu}.
So the bundle gerbe $J$ is defined as  a hermitian line bundle over
the fibre product ${\mathcal A}_0^{[2]}$. This
fibre product can be identified with all triples $(A, \l, \l')$
where neither $\l$ nor $\l'$ are in the spectrum of $D_A$. The fibre
of $J$ over $(A, \l, \l')$ is just $DET_{\l\l'}$.  For this to
be a bundle gerbe we need a product which in this case is a linear
isomorphism
$$
DET_{\l\l'} \otimes DET_{\l'\l''} =   DET_{\l\l''}.
$$
But such a linear isomorphism
 is a simple consequence of the definition of
$DET_{\l\l'} $ and the fact that taking top exterior
powers is multiplicative for direct sums.

Let $\pi \colon {\mathcal A}_0 \to \mathcal A$
be the projection and
 $p \colon {\mathcal A} \to {\mathcal A}/{\mathcal G}_{e}$ be the quotient
by the gauge action.
It was shown in \cite{CaMu1} that the line bundle $DET$ on ${\mathcal A}_0$
satisfies
$J =  \delta(DET)$.  Here $\delta(DET) =  \pi_1^*(DET)^*\otimes \pi_2^*(DET)$
where $\pi_i \colon {\mathcal A}_0^{[2]} \to {\mathcal A}_0$ are the
projections,
$$
\pi_1((A, \l, \l')) =  (A,  \l), \quad
\pi_2((A, \l, \l')) =  (A,  \l').
$$
In other words $J= \delta(DET)$ is equivalent to
$$
DET_{\l\l'} =  DET_{\l}^*\otimes DET_{\l'}
$$
which is equivalent to equation \ref{det_eq}.

The fibering ${\mathcal A}_0 \to \mathcal A$ has, over each open set
$U_\lambda$
a canonical section $A \mapsto (A, \lambda)$.
These enable us to suppress the geometry of the submersion
and the bundle gerbe $J$ becomes the line bundle $DET_{\lambda\l'}$
over the intersection $U_{\l\l'}$ and its triviality amounts
to the fact that we have the line bundle $DET_\l $ over $U_\l$
and over intersections we have the identifications
$$
DET_{\l\l'} =  DET_{\l}^*\otimes DET_{\l'}.
$$

We denote the Chern class of $DET_{\l\l'}$ by $\theta_2^{[2]}$.
Note that these bundles descend to
bundles over $V_{\l\l'}= \pi(U_{\l\l'})\subset {\mathcal A}/{\mathcal G}_e.$
Therefore, the forms $\theta_2^{\l\l'}= \theta_2^{\l}-\theta_2^{\l'}$
on $U_{\l\l'}$ (where $\theta_2^{\l}$ is the 2-form giving the
curvature of $DET_{\l}$) are equivalent (in cohomology) to forms
which descend to closed 2-forms $\phi_2^{\l\l'}$ on $V_{\l\l'}.$
The following result is established in \cite{CaMiMu}

\begin{theorem}\cite{CaMiMu}
The family of closed 2-forms $\phi_2^{\l\l'}$ on $V_{\l\l'}$
determines a representative for the Dixmier-Douady class $\omega$
of the bundle gerbe $J/{\mathcal G}_e$.
In addition, noting that $\delta(DET)= J$, the connection with the
 Faddeev-Mickel\-sson cocycle on the Lie algebra of the gauge group
is simply that it is cohomologous to the negative of the cocycle
defined by the curvature $F_{DET}$ of the line bundle $DET$ along
gauge orbits.
\end{theorem}

To obtain the Dixmier-Douady class as a characteristic class
we recall that
in the case of even dimensional manifolds, Atiyah and Singer \cite{AtSi}
gave a construction of `anomalies' in terms of characteristic classes.
In the present case of odd dimensional manifolds
 a similar procedure yields the Dixmier-Douady class.

We begin with the observation that
given a closed integral form $\Omega$ of degree $p$ on a product manifold
$M\times X$ (dim$M= d=2n+1$ and dim$X= k$) we obtain a closed integral form
on
$X,$ of degree $p-d,$ as
$$\Omega_X =  \int_M \Omega.$$
If now $A$ is any  Lie algebra valued connection on
the product $M\times X$ and $F$ is the corresponding curvature we can con=
struct
the Chern form $c_{2n} =  c_{2n}(F)$ as a polynomial in $F.$ Apply this
to the
connection $A$ defined by Atiyah and Singer, [AtSi], [DoKr p. 196], in the
case when $X= {\mathcal A}/{\mathcal G}_e.$

First pull back the forms to $M\times \mathcal A$.
The Atiyah-Singer connection on
$M\times X$ becomes a globally defined Lie algebra valued 1-form $\hat A$
 on
$M\times \mathcal A$. Let $\hat F$ be the curvature form
determined
by $\hat A.$ We showed in \cite{CaMiMu} that
$$\int_{S^3} \Omega_X = \int_{M\times D^3} c_{2n}(\hat F).$$
where the disk $D^3$ is the pullback to $\mathcal A$ of $S^3\subset
{\mathcal A}
/{\mathcal G}_e.$
But the integral of the Chern form over a manifold with a boundary (when
the
potential is globally defined) is equal to the integral
of the Chern-Simons action:
$$\int_{M\times \partial D^3} CS_{2n-1}(\hat A).$$
Along gauge directions the form $\hat A$ is particularly simple
so for example
when $M= S^1$ and $2n=4$ we get  (here $S^2=\partial D^3$)
$$\int_{S^3} \Omega_X =\int CS_3(\hat A) =  \frac{1}{24\pi^2}\int_{S^1\times
S^2}
\text{tr} (dg g^{-1})^3$$
where $g=g(x,z),$ $z\in S^2,$ is a family of gauge transformations
relating the vector potentials on the boundary $S^2=\partial D^3.$
Similar results hold in higher dimensions.
\begin{theorem}\cite{CaMiMu}
The class $\Omega_X$ is a representative for  the Dixmier-Dou\-ady class
of the
bundle gerbe $J/{\mathcal G}_e$.
\end{theorem}

\section{$\U$ bundles and string structures}
\label{Ures}
Let $H= H_+\oplus H_-$ be a polarization of a Hilbert space $H$ into a
pair
of closed infinite-dimensional subspaces. We denote by
${\mathcal U}_{res}$ the
restricted unitary group consisting of unitary operators in $H$ such that
the off-diagonal blocks are Hilbert-Schmidt operators.
In a recent preprint \cite{CaCrMu}
we described in some detail results about the Dixmier-Douady class
arising from the problem of lifting
principal ${\mathcal U}_{res}$ bundles
to principal $\hat\U$ bundles. Here $\hat{\mathcal U}_{res}$ is a central
extension of $\U,$ [PrSe].
We now summarise the results in \cite{CaCrMu}.

\begin{theorem}
There is an imbedding of the smooth loop group $LG$ of a compact Lie group
$G$ in  ${\mathcal U}_{res}$ which extends to give an
imbedding of the canonical central extension $\widehat{LG}$
in $\hat{\mathcal U}_{res}$. Under this imbedding
the obstruction to the existence
of a  string structure (in the sense of Killingback \cite{Kil}: a
$\widehat{LG}$
principal bundle) on the loop space of a manifold $M$
may be identified with the Dixmier-Douady class
of the lifting bundle gerbe of the corresponding
principal ${\mathcal U}_{res}$ bundle.
\end{theorem}

A different approach to the question of the existence of
string structures
is due to \cite{Mc} and exploits Brylinski's point of view
whereas in \cite{CaMu} the problem
is solved using the classifying map of the bundle over the loop space of
$M$.

There is also an imbedding of  ${\mathcal U}_{res}$
in the projective unitary group of the skew symmetric Fock space (determined
by the polarization $H= H_+\oplus H_-,$ the 'Dirac sea' construction),
[PrSe].
Under conditions on the
underlying manifold $M$ this imbedding enables us to
establish a relationship between the
Dixmier-Douady class of a bundle gerbe over
$M$ determined by a  ${\mathcal U}_{res}$
bundle and the second Chern class of an associated principal
projective unitary group bundle over the suspension of $M$.

In this section we
describe the field theory examples which motivated the proving of the
previous results.

Let $\Gr$ be the space of all closed infinite-dimensional subspaces of
$H$ with the topology determined by operator norm topology for the
associated projections. We may think of $\Gr$ as the homogeneous space
$$ U(H)/( U(H_+) \times U(H_-)).$$
Here all the groups are contractible
(in the operator norm topology) and therefore there is a continuous
section $\Gr \to U(H),$ that is, for $W\in \Gr$ we may choose a $g_W \in
U(H)$ which depends continuously on $W,$ such that $W=  g_W \cdot H_+.$

The example we shall study below comes from a quantization of a family
of Dirac operators $D_A$ parametrized by smooth (static) vector potentials
 $A.$
In the following we shall use the notations in section 4.

Choose a real number $\l$ such that $D=D_A -\l$ is invertible.
The set of bounded operators $X$ such that $||X|| < ||1/D||^{-1}$ is an
open set $V$ containing $0$ and the function $X\mapsto |D +X|^{-1}(D_0+X)$ is
continuous in the operator norm of $X;$ this is seen using the
converging geometric series $(D +X)^{-1}=D^{-1} -D^{-1} X
D^{-1} + \dots.$ Since the operator norm of the
interaction $A$ depends continuously on the components $A_i$ of the
vector potential (with respect to the infinite family of Sobolev norms
on $\mathcal A$) we can conclude
that $A\mapsto \epsilon_{A,\l}=  (D_A-\l)|D_A-\l|^{-1}$ is continuous.
Thus also the spectral projections $P_{\pm}(A,\l) =  \frac12 (1 \pm
\epsilon_{A,\l})
$ are continuous and $H_+(A,\l)=  P_+(A,\l) H \in \Gr$ depends
continuously on $A\in U_{\l}$. On the other hand,
we know that there is a section $\Gr \to U(H)$ and therefore we may choose
a continuous function $A\mapsto g_{\lambda}(A) \in U(H)$ such that
$H_+(A,\lambda) =  g_{\lambda}(A) \cdot H_+.$  We shall show that these
define
transition functions, $g_{\l\l'}(A)= g_{\l}(A)^{-1} g_{\l'}(A),$ for a
principal ${\mathcal U}_{res}$ bundle $P$ over $\mathcal A.$ By construction,
these satisfy
the cocycle property required for transition functions so the only thing
which
remains is to prove continuity with respect to the topology of
${\mathcal U}_{res}.$

The topology of $\U$ is defined by the operator norm topology on the diagonal
blocks (with respect to the energy polarization $H_+\oplus H_-$ fixed by
a
'free' Dirac operator $D_{A_0}$ without zero modes) and by
Hilbert-Schmidt norm topology on the off-diagonal blocks. As before,
$P_{\pm}= P_{\pm}(A_0,0)$ and we set $\epsilon=  P_+ - P_-.$
We already know that the $g_{\l\l'}$'s (assume e.g. that $\l <\l'$)
are continuous with respect to the
operator norm topology and we need only show that the off-diagonal
blocks $[\epsilon,g_{\l\l'}]$ are continuous in the Hilbert-Schmidt topology.
Let us concentrate on the upper right block $K_{+-}= P_+ g_{\l\l'} P_-.$

Multiplying from the left by $g_{\l}$ and from the right by $g_{\l'}^{-1}$
and using the fact that Hilbert-Schmidt operators form an operator ideal
with $||gK||_2 \leq ||g||\cdot ||K||_2$ we conclude that $K_{+-}$ is continuous
in the Hilbert-Schmidt norm if and only if
$$g_{\l} P_+ g_{\l}^{-1} g_{\l'} P_- g_{\l'}^{-1}$$
is a continuous function of $A$ in the Hilbert-Schmidt norm. Now the product
of the first three factors in the above expression gives $P_+(A,\l)$
whereas the product of the last three factors is $P_-(A,\l').$ But
$P_+(A,\l) P_-(A,\l')$ is the spectral projection
$P(\lambda,\lambda')$ to the finite-dimensional
spectral subspace corresponding to the interval $[\l,\l'].$ On the other
hand, the dimension of this subspace is fixed over $U_{\l\l'}$ and
therefore the Hilbert-Schmidt norm of the projection, which is the square
root of its rank, is continuous. Furthermore, since $P(\l,\l')$ is continuous
in the operator norm and it has a fixed finite rank it is also continuous
in the Hilbert-Schmidt norm.

We denote by $\Gr_{res}$ the restricted Grassmannian, defined as the orbit
${\mathcal U}_{res} \cdot H_+$ in $\Gr.$
The fiber $P_A$ at $A\in \mathcal A$ can be thought of as the set of all
unitary
operators $T:H\to H$ such that $T^{-1}( H_+(A,\l))$ (for any $\l$) is in
$\Gr_{res}.$ This is because $g_{\l}(A)$ provides such an operator for
any $A\in U_{\l}$ and any two such operators differ only by a
right
multiplication by an element of $\U$

Being a principal bundle over a contractible parameter space, $P\to
\mathcal A$ is trivial.  We choose a global trivialization $A\mapsto T_A.$

On any $U_{\lambda}$ the function $A\to  P_+(A,\lambda)$ is
continuous and $$T_A^{-1} P_+(A,\lambda) T_A \in \Gr_{res}.$$
 Over $\Gr_{res}$ there is a
canonical determinant bundle $DET_{res}.$ The action of $\U$ on
$\Gr_{res}$ lifts to an action of $\hat\U$ on $DET_{res},$ [PrSe].

Using the maps $A\to P_+(A,\lambda)$
we can pull back the determinant bundle $DET_{res}$
over $\Gr_{res}$ to form local determinant bundles $DET_{\lambda}$ over
$U_{\lambda}.$ This family is the right one for discussing the
gerbes over $\mathcal A$ and ${\mathcal A}/{\mathcal G}_e.$
The reason is that the class
of the bundle gerbe is completely determined by the line bundles $DET_{
\lambda\lambda'}$ over $U_{\lambda\lambda'}.$

On the restricted
Grassmannian we obtain an isomorphism between the fibers $DET_{res}(W)$ and
$DET_{res}(W')
$, where $W'\subset W$ are  points in $\Gr_{res};$  the isomorphism is
determined by a choice of
basis $\{v_1,\dots, v_n\}$ in $W \cap W^{\prime^{\perp}}$ as follows.
 Recalling from [PrSe] that
an element in $DET_{res}(W)$ is represented by the so-called admissible basis
$\{w_1,w_2,\dots\},$ modulo unitary rotations with determinant equal to one,

the isomorphism is simply $\{w_1,w_2,\dots,\} \mapsto \{v_1,\dots,v_n,w_1,
w_2, \dots\}.$ In particular, we apply this when $W,W'$ are
the points obtained by mapping $H_+(A,\lambda)$ and
$H_+(A,\lambda')$ to $\Gr_{res}$ using ${T_A}^{-1}.$
Now the vectors
${T_A}^{-1} v_i$ span a basis in the subspace corresponding to the interval
$[\lambda,
\lambda']$ in the spectrum of $D_A$ and thus they define an element in
$DET_{\lambda\lambda'}$ in our earlier construction and the basis can be
viewed
as an isomorphism between $DET_{\lambda}$ and $DET_{\lambda'}.$

Next we consider the trivial bundles ${\mathcal A} \times \U$ and
${\mathcal A}\times \hat \U$ over $\mathcal A$. The gauge group
$\mathcal G$ acts
in the former as follows. Define $\omega(g;A)=  T_{g\cdot A}^{-1} g T_A$ This
function takes values in $\U$ and is a 1-cocycle by construction, [Mi3],
$$\omega(gg';A) = \omega(g;g'\cdot A) \omega(g';A).$$
Thus the gauge group acts through $g\cdot (A, S)=  (g\cdot A, \omega(g;A)S)$
in ${\mathcal A}\times \U.$

Since $\omega$ takes values in $\U$ the same construction which gives
the lifting of the $\U$ action on $\Gr_{res}$ to a $\hat\U$
action on $DET_{res}$
gives also an action of an extension $\hat{\mathcal G}$ in
${\mathcal A}\times\hat\U$
and in ${\mathcal A} \times DET_{res}.$
The pull-back with respect to the conjugation by $T_A$'s of the latter action
defines an action of $\hat {\mathcal G}$ on the local determinant bundles
 $DET_{\lambda}.$
Next  we observe that the natural action (without center)
$v_i \mapsto gv_i$ in the line $DET_{\lambda\lambda'}$ intertwines between
the action of the group extension in the lines
$DET_{\lambda},DET_{\lambda'}$
parametrized by potentials $g\cdot A$ on the gauge orbit.
This follows from the corresponding property of the determinant bundle
over $\Gr_{res}$ (by pushing forward by $T_A$): An element  $\hat g\in
\hat\U$ acts
on $w= \{w_1,w_2,\dots,\} \in DET_{res}(W)$ as $w_i\mapsto \sum_j gw_j =
q_{ji},$
where the basis rotation $q$ is needed in order to recover a basis in the
admissible set, \cite{PrSe}. The same element $\hat g$ acts then on the
basis
$w'= w\cup v$ extending the action on $w$ by sending $v_i$ to $gv_i.$

The intertwining property of the natural action on $DET_{\lambda\lambda'}$
is exactly what was needed in the definition of the action of $\hat
{\mathcal G}$ in the Fock bundle over $\mathcal A$. On the other hand,
the obstruction
to pushing the Fock bundle over ${\mathcal A}/ {\mathcal G}_e$
was precisely the class of
the extension $\hat {\mathcal G} \to {\mathcal G}.$ Thus we have

\begin{theorem} The obstruction to pushing forward the trivial bundle
${\mathcal A}
\times \hat \U$ to a bundle over the quotient ${\mathcal A}/{\mathcal G}_e,$
with the action of $\hat{\mathcal G}$ coming from the
$\U$ valued cocycle $\omega,$ is the Dixmier-Douady class of the
Fock bundle. \end{theorem}

It is clear from the above discussion that we may view the Fock bundle
over $\mathcal A$ as an associated bundle to the principal bundle
${\mathcal A}\times \hat \U$ defined by the representation of $\hat \U$ in
the Fock space of free fermions.

\noindent\bf Example \rm Let us take a very concrete example for the discussion
above. Let $G= SU(2)$ and the physical space $M= S^1.$
 Now ${\mathcal A}/{\mathcal G}_e$
is simply equal to $G$ since the gauge class of the connection in one
dimension is uniquely given by the holonomy around the circle.
Because topologically $SU(2)$ is just the unit sphere $S^3$ any principal
bundle over $G$ is described by its transition function on the equator
$S^2.$ In case of a $\U$ bundle we thus need a map $\phi:S^2 \to
\U$ to fix the bundle and the equivalence class of the bundle is
determined by the homotopy class of $\phi.$ The topology of $\U$ is
known: it consists of connected components labelled by the Fredholm index
of $P_+ g P_+,$ it is simply connected and so the second homotopy is given
by $H^2(\U, \Bbb Z) = \Bbb Z.$ Thus the equivalence class of a
principal $\U$ bundle over $S^3 \equiv {\mathcal A}/{\mathcal G}_e$
 is given by the
index of the map $\phi.$

The principal ${\mathcal G}_e$ bundle
${\mathcal A} \to {\mathcal A}/{\mathcal G}_e$ is defined by a transition
function $\xi: S^2 \to {\mathcal G}_e.$
This is determined as follows.
Since the total space is contractible, we actually have here a universal
${\mathcal G}_e$ bundle over $S^3.$
Thus the transition function $\xi$ is the
generator in $\pi_2({\mathcal G}_e).$
Such a map can be explicitly constructed.
Any point $Z$ on the equator $S^2\subset S^3$ determines a unique half-circle
connecting the antipodes $\pm 1.$
We define $g_Z: S^1 \to SU(2)$ by first following the great circle through
a fixed reference point $Z_0$ on the equator,
as a smooth function of a parameter $0\leq x\leq \pi-\delta$ (where $\delta$
is a small positive
constant), from the point $+1$ to the antipode $-1.$ For parameters
$\pi-\delta < x <\pi +\delta$ we let $g_Z(x)$ to be constant,
for $\pi+\delta \leq x\leq 2\pi-\delta$ the loop continues from $-1$ to
$+1$
through the point $Z$ on the equator,
and finally for $2\pi-\delta \leq x \leq 2\pi$ it is constant.
It is easy to see that the set of
smooth loops so obtained covers $S^3$ exactly once and therefore gives
a map $g:S^2\to {\mathcal G}$ of index one.

Any element of $\mathcal G$
is represented as an element of ${\mathcal U}_{res}$
through pointwise multiplication
on the fermion field in $H.$ Thus by this embedding we get directly the
transition function $\phi$ for the ${\mathcal U}_{res}$
bundle over ${\mathcal A}/{\mathcal G}_e.$

The index of the map $\xi$ can also be checked using the WZWN action,
$$\text{ind}\,\xi =  \frac{1}{24\pi^2} \int_{S^2\times S^1} \TR (g^{-1}dg)^3$$
and in the fundamental representation of $G= SU(2)$ this gives ind$\,\xi= 1.$
For chiral fermions on the circle in the fundamental representation of
$G$ this is the same as the index of the map $\phi:S^2 \to \U.$
This latter
index is evaluated by pulling back the curvature form $c$ on $\U$ to $S^2$
and then integrating over $S^2.$ The curvature is defined by the same formula
as the canonical central extension of the Lie algebra of $\U.$ Identifying
left-invariant vector fields on the group manifold as elements in the Lie
algebra we have
$$c(X,Y)= \frac14\TR\epsilon[\epsilon,X][\epsilon,Y],$$
where $\epsilon$ is now defined by the polarization to nonnegative and negative
Fourier modes.
Note that this curvature on $\U$ is the generator of $H^2(\U,
\Bbb Z).$

The Dixmier-Douady class in our example,
 as a de Rham class in $H^3({\mathcal A}/
{\mathcal G}_e),$ is simply the normalized volume form on $S^3.$ This is
because the third cohomology group of $S^3$ is one-dimensional  and the
Dixmier-Douady class was constructed starting from the universal bundle
${\mathcal A}\to
S^3.$

\section{Global anomalies}

\subsection{Bundle gerbes with other structure group}

There is  no particular  reason to restrict attention to
$\C^\times$ as the structure group for bundle gerbes.
 If $Z$ is any abelian topological
 group there is a theory of $Z$ bundle gerbes
obtained by
replacing $\C^\times$ by
$Z$ throughout.  We need $Z$ abelian in order to
take tensor products of $Z$ bundles - (this does not work if $Z$ is not
abelian).
Brylinski calls these: gerbes with `band' $\underline Z$ (where $\underline Z$
is the sheaf of smooth functions into $Z$).  In such a theory the
Dixmier-Douady class
is in $H^2(M, \underline Z)$ because the isomorphism $H^2(M, \C^\times)
= H^3(M, \Z)$ is generally not available.

In particular if $Z$ is a subgroup of $\C^\times$ one may think of
a $Z$ bundle gerbe as a special ordinary bundle gerbe. It is one
where the $\C^\times$ bundle $P \to Y^{[2]}$ has a reduction to $Z$ and
that reduction is preserved by the bundle gerbe product. In such a case
the Cech cocycle which is {\it a priori} in $H^2(M, \C^\times)$
 naturally ends up in
$H^2(M, Z).$

If there is a central extension
$$Z \to \hat G \to G$$
and a $G$ bundle
$P \to M$ there is a lifting $Z$ bundle gerbe whose Dixmier-Douady class
is the
obstruction to lifting $G$ to $\hat G.$ This may be seen by noting
its construction. If $P \to M$ is a $G$ bundle there is a map
    $$s\colon P^{[2]} \to G$$
defined by $ps(p,q) = q$. Then the $Z$ bundle gerbe, as a $Z$ bundle over
$P^{[2]}$ is
just the pull-back of $\hat G \to G$ under $s$. Hence any special properties
of $\hat G \to G$ are inherited by the bundle gerbe.

There is also a theory of flat bundle gerbes.
  If $L \to M$ is a flat line bundle with connection
$\nabla$ then it can be represented locally by transition functions $g_{ab}$
that are locally constant. Hence its Chern class is in $H^1(M, \C^\times)$
rather than $H^1(M, \underline \C^\times)$. In fact one can show that
flat line bundles are classified by $H^1(M, \C^\times)$.

Similarly stable isomorphism classes of flat gerbes with connection and
curving are classified by $H^2(M, \C^\times)$.
Particular examples of these can be obtained by looking at $H^2(M, \Z_n)$
where $\Z_n$ is the cyclic subgroup of $U(1)$.  These are flat line bundles
whose Chern class is $n$ torsion.

One can also realise the
Dixmier-Douady class as an element of $H^3(M, \Z)$
when $Z$ is say $\Z_n$.  To see this consider the following commuting
diagram of
sheaves of groups
$$
\begin{array}{ccccccc}
0 \to &\Z &\to &\underline{\C} &\to  &{\underline{\C}}^\times &\to 0 \\
      &\uparrow &  &\uparrow  & &\uparrow & \\
0 \to &\Z &\to  &\frac{1}{n} \Z & \to &\Z_n& \to 0
\end{array}
  $$
Here the first vertical arrow is an equality, and the second and
third are inclusions.
The coboundary map for the lower short exact sequence
induces the so-called Bockstein map $\beta^*: H^2(M, \Z_n)\to H^3(M, \Z)$
whose image consists of $n$ torsion classes in $H^3(M, \Z)$.

\subsection{The framework for the examples}

The notation is as in Section 4.
Thus $H$ is the tensor product of the Hilbert space of square-integrable
spinor fields on a
compact Riemannian spin manifold $M$ and a finite-dimensional inner
 product space
$V.$ We assume that an action of a compact Lie group $G$ on $V$ is given.
 This
gives a natural action of ${\mathcal G}= Map(M,G)$ on $H.$  We have a
polarization
$H=H_+\oplus H_-$ corresponding to the splitting of the spectrum of the
 Dirac
operator $D=D_{A_0}$ on $M$ to nonnegative and negative parts.

The Lie algebra of $\U$ has a central extension defined by the cocycle
$$c(X,Y)= \frac14 \TR\epsilon[\epsilon,X][\epsilon,Y]$$ and
the corresponding group extension $\Uhat$ is a topologically nontrivial
circle bundle over $\U$. This bundle has a natural connection defined by
the 1-form $\theta = \text{pr}_c g^{-1} dg,$ the central projection of
the
Maurer-Cartan form on $\Uhat.$ The curvature $\O$ of this form is
left-invariant
and at $g=1$ it is given by the 2-cocycle $c.$ The curvature is \it integral,
\rm its integral over a closed surface is an integer.

Starting from the Lie algebra central extension (or curvature form) one
can construct $\Uhat$ as follows. Consider the set $\mathcal P$ of smooth
 paths
$g(t) \in \U,$
$0\leq t\leq 1,$ with $g(0) =1$ and $g(t)=g\in \U$ Define and equivalence
relation in ${\mathcal  P}\times S^1$ by $(g_1(\cdot),\lambda) \sim
(g_2(\cdot),\mu)$
if $g_1(1)=g_2(1)$ and $\mu=\lambda \cdot \exp(2\pi i \int_D \Omega),$ where
$D$ is any surface in $\U$ such that the boundary of $D$ is the union of
the paths $g_1$ and $g_2.$ Define a product in ${\mathcal P}\times S^1$ as
$(g_1(\cdot),\lambda)\cdot(g_2(\cdot),\mu)=(g_3(\cdot),\lambda\mu),$ where
$g_3(t)=g_1(t)g_2(t).$ Then $\Uhat =({\mathcal P}\times S^1)/\sim.$

As before, we construct the 1-cocycle $\omega(g;A) = T_{g\cdot A}^{-1}g T_A$
with values in $\U.$
The obstruction to pushing forward the bundle ${\mathcal A}\times
{\mathcal F}$ of
Fock spaces over $\mathcal A$ to a bundle $({\mathcal A}\times
{\mathcal F})/{\mathcal G}_e$ over
${\mathcal A}/{\mathcal G}_e$ is the obstruction to lifting the cocycle
$\omega$ to a
cocycle $\hat\omega$ with values in $\Uhat.$

We have earlier discussed the local part of this obstruction. The local
obstruction is due to the fact that the pull-back with respect to the
map $g\mapsto \omega(g;A)$ of the circle bundle $\Uhat$ over $\U$ might
be nontrivial. (The gauge parameter $A$ is irrelevant in this context bec=
ause
$\mathcal A$ is an affine space and so uninteresting for the problem of
nontriviality of bundles over ${\mathcal G}_e\times {\mathcal A}.$ ) The
circle bundles
are classified by the Chern class, given as a (cohomology class of) 2-form.
This was related, via families index theorem, to the Dixmier class on
${\mathcal A}/{\mathcal G}_e. $

 Now we shall assume that the local obstruction vanishes, i.e. the
restriction of the curvature form $\Omega$ to the submanifold $\{\omega(g;A)
| g\in {\mathcal G}_e\} \subset \U$ vanishes (for all $A\in
{\mathcal A}.$)

\subsection{The case $G=SU(2)$}
This is the original case considered by Witten in even dimensions, [Wi1].
In our situation
 the dimension of $M=S^3$ is three and then the
curvature form of the local determinant bundles along gauge orbits is
$$ \frac{i}{24\pi^3} \int_M \TR A [dX,dY] \equiv 0.$$
This follows from $\TR X(YZ +ZY)\equiv 0$ in the Lie algebra of $SU(2).$
On the other and, as we have seen, the curvature of the determinant bundles
gives directly the 2-cocycle of the Lie algebra extension arising from the
action in Fock spaces. Even if the
 local obstruction vanishes there can be a finite torsion
obstruction for lifting the cocycle $\omega$ to $\hat \omega.$ This
obstruction can arise only if $\pi_1({\mathcal G}_e) \neq 0.$ The reason
for this
is understood using the construction in (2). If $g_1(t)$ and $g_2(t)$ are
two paths with the same end points then $g_1$ is always homotopic to $g_2$
(with end points fixed) if $\pi_1{\mathcal G}_e=0.$ But now the curvature
$\Omega$
vanishes along ${\mathcal G}_e$ and thus we have a lift $\omega(g;A) \mapsto
\hat\omega(g;A)= [(\omega(g(\cdot);A),1)]$ where $g(t)$ is any path joining
$g$ to $1$ in ${\mathcal G}_e$ and the outer brackets denote equivalence
classes
modulo the relation defined in 6.2.
If $\pi_1({\mathcal G}_e)\neq 0$ we have to
examine further the existence of the obstruction. Note that in the case of
$G=SU(2)$ and dimension three, $\pi_1({\mathcal G}_e)=\mathbb Z_2.$

 There is a homomorphism of $\pi_1({\mathcal G}_e)$ to $S^1$ defined by
$\phi(g(\cdot)) = \exp(2\pi i \int_D \Omega),$ where $D\subset\U$ is
any surface with boundary curve $\omega(g(t);A).$ Since $\mathcal A$ is
connected the equivalence class of this discrete group representation
cannot depend on the continuous parameter $A$ and we can fix $A=0,$ for
example. The torsion obstruction is then the potential nontriviality
of this representation.

 In order to determine the relevant representation we have to compute
$\int_{D_i} \Omega$ for a
 set of generators $g_i(t)$ of $\pi_1({\mathcal G}_e)$
with $\partial D_i =\omega(g_i(\cdot);0)\subset \U.$ In the case of
$G=SU(2)$ in the defining representation and dimension $=3$ this is
particularly simple. We use a trick
due to Witten, [Wi2]. Embed $SU(2) \subset SU(3)$ and use the fact that
$\pi_4(SU(3))=0$ and on the other hand, $\pi_1(Map(M,SU(3))) =\pi_4(SU(3)).$

Correspondingly, we extend the number of (internal) Dirac field components
from 2 to 3 and we have, in a self-explanatory notation, $\U^{(2)} \subset
\U^{(3)}.$ The restriction of the curvature form on $\U^{(3)}$ to the
subgroup gives the curvature on the former group. Since $\pi_1({\mathcal
G}_3)=0,$
we can choose $D\subset Map(M,SU(3))$ such that the boundary of $D$ gives
 the
generator of $\pi_1({\mathcal G}_2)=\mathbb Z_2.$

For a given $A\in \mathcal A$ the pull-back of $\Omega$ with respect to the
map $g\mapsto \omega(g;A)$ is equal to the 2-form
\begin{eqnarray*}
 \Phi_A(X,Y;g)= & \Omega\left(\frac{d}{dt} \omega(g\cdot e^{tX};A)|_{t=
0},
\frac{d}{ds}\omega(g\cdot e^{sY};A)|_{s=0}\right)\\
&= \Omega\left( \omega(g;A) \frac{d}{dt} \omega(e^{tX};A^g)|_{t=0},\omega(g;A)
\frac{d}{ds}\omega(e^{sY};A^g)|_{s=0} \right)\\
&=\Omega\left(\frac{d}{dt}\omega(e^{tX};A^g)|_{t=0},
\frac{d}{ds}\omega(e^{sY}; A^g)|_
{s=0}\right)
\equiv c(X,Y;A^g)
\end{eqnarray*}
where we have used the left-invariance of the form $\Omega$ and $c(X,Y;A)$
is the Schwinger term induced by the cocycle $\omega$ and the central
extension of $\U.$

 For $G=SU(2)$ and $d=3$ the above formula gives
$$\Phi_A(X,Y)  \sim \int_M \TR A^g [dX,dY] \equiv 0.$$
If $D\subset Map(M,G)$ is a disk (and the dimension of $M$ is 3)
parametrized
by real paramaters $t,s$ then
$$\int_D \Phi_A = \frac{i}{24\pi^3}
\int_D \int_M \TR A^g [d(g^{-1}\partial_t g), d(g^{-1} \partial_s g)].$$
In particular, at $A=0$ the result is
$$\int_D \Phi_A = \frac{i}{480\pi^3} \int_{D\times M} (g^{-1} dg)^5$$
provided that we can ignore boundary terms in integrations by parts; this
 is
the case if at the boundary of the disk $g(t,s) \in Map(M,SU(2)).$
In this case the last integral has been computed in \cite{Wi2}; the result is
$1/2$ mod integers if the boundary circle is represents the nonzero element
in $\pi_4(SU(2)),$ otherwise the integral is zero mod integers. Since
$\exp(2\pi i \int_D \phi)$ is the factor appearing in the definition of the
extension of the group of gauge transformations in the Fock spaces, this
result shows that the double cover of $Map(M,SU(2))$ is represented
nontrivially
and therefore obstructing the lifting of the cocycle $\omega$ to a quantum
extension $\hat\omega.$

The global $SU(2)$ anomaly in the bundle of Fock
vacua can also be analyzed in
terms of the spectral flow of a family of Dirac hamiltonians, \cite{NeAl}.
The $\Bbb Z_2$ extension of the gauge group $Map(M,SU(2))$ has been used
for
deriving a boson-fermion correspondence in four space-time dimensions,
\cite{Mi2}.

\subsection{A general analysis}
The idea of the $SU(2)$
example should work for any simple group $G$ in any dimension.
Given a complex $k-$ dimensional representation of $G$ acting on the
spinor components we extend the number of components to a large value
$N$ and think of $G$ as a subgroup of $G_{\infty}=SU(N),$ \cite{ElNa}.
If the dimension of
$M$ is $d=2n+1$ then $\pi_{d+1}(SU(N))=0$ for
large enough $N.$ So given a loop $\gamma$ in $Map(M,G)$ we can find a
disk $D$ in $Map(M,G_{\infty})$ such that $\partial D =\gamma.$ Finally,
checking
the nontriviality of the obstruction, comes up to evaluating the integral
$$\left(\frac{-i}{2\pi}\right)^{n+2} \frac{(n+1)!}{(d+2)!}
\int_{D\times M} \TR (g^{-1} dg)^{d+2}$$
and checking whether it is zero mod integers.

In order that there could be a non-trivial global anomaly we note that we
 need
to be in a situation where
 $\pi_1({\mathcal G}_e)$ is non-trivial, say equal to $\Z_n$
for some integer $n$, and there is no local anomaly. Then one
 has a central extension
$$\Z_n\to \hat{\mathcal G}_e\to {\mathcal G}_e.   \eqno(6.1)$$
The Cech 1-cocycle arising from this extension
takes values in $\Z_n\subset U(1)$.
Using the usual exact sequence
$$\Z\to R\to U(1)$$
to change coefficients
one sees that this
gives the Chern class as a torsion element of the Cech cohomology
group $H^2({\mathcal G}_e, \Z)$.

We may consider the corresponding  lifting bundle gerbe for
the principle ${\mathcal G}_e$  bundle
$${\mathcal G}_e\to {\mathcal A}\to {\mathcal A}/{\mathcal G}_e.$$
This lifting bundle gerbe has Dixmier-Douady class in
$H^2({\mathcal A}/{\mathcal G}_e, \underline{U(1)})$
which, using the exact sequence
$$\Z \to R\to U(1)$$
is represented by a torsion class in $H^3({\mathcal A}/{\mathcal G}_e,
\Z)$.
The argument of Theorem 4.1 of \cite{CaCrMu} shows that
the Dixmier-Douady class is the transgression of the Chern class of
the extension (6.1) as an element of
$H^2( {\mathcal A}/{\mathcal G}_e, \Z)$.

In the $SU(2)$ example we have a reduction of the (local) determinant bundles
along gauge orbits in $\mathcal A$ to $\Z_2$ bundles. On ${\mathcal
A}/{\mathcal G}_e$ this
corresponds to trying to lift the system of local $\hat{\mathcal G}_e$
(where $\hat{\mathcal G}_e$ is the $\Z_2$ extension of the group of gauge
transformations) to a global $\hat{\mathcal G}_e$ bundle.
The obstruction is the Dixmier-Douady class: our torsion element
in $H^3({\mathcal A}/{\mathcal G}_e)$.

It is of interest to have a practical method for
determining when the global Hamiltonian anomalies
are non-trivial.
There is a method for finding
 the extension $\hat{\mathcal G}_e$ of ${\mathcal G}_e$
which acts on $DET_\lambda$ for each $\lambda$.
The map which sends $g\in\hat{\mathcal G}_e$
to $g^k$ is a homomorphism onto  ${\mathcal G}_e$
for sufficiently large $k$. Choose the smallest such $k$.
Then as  $\hat{\mathcal G}_e$ acts on $DET_\lambda$
so ${\mathcal G}_e$ acts on  $(DET_\lambda)^k$.
Thus the bundle gerbe given locally by
$$(DET_{\lambda\lambda^\prime})^k=
(DET_{\lambda}^*)^k \otimes (DET_{\lambda^\prime})^k$$
admits an action of ${\mathcal G}_e$ on each factor in the tensor product
and so descends  to a trivial bundle gerbe over ${\mathcal A}/{\mathcal G}_e$.
Finally, finding $k$ may be done using the Witten method.

As above we have a compact Lie group $G$
with say $\pi_{d+1}(G)$ torsion, say $\Z_n$, for $d$ odd.
Then the subgroup ${\mathcal G}_e$ of based
gauge transformations in $Map(S^{d}, G)$ has $\pi_1{\mathcal G}_e=\Z_n$.
For large enough $N$ we have $\pi_{d+2}(G_\infty)=\Bbb Z$ in addition to
$\pi_{d+1}(G_\infty)=0.$
We assume that the local Hamiltonain anomaly for the pair $S^{d-1},G$
vanishes and we wish to know if there exists a global anomaly.
The imbedding into $G_\infty$ enables us to exploit Witten's trick.
Consider part of the homotopy long exact sequence for the
fibration
$$G\rightarrow G_\infty\rightarrow X$$
where $X$ denotes the quotient space $G_{\infty}/G$:
$$
\dots\rightarrow\pi_{d+2}(G_\infty)\rightarrow\pi_{d+2}(X)\rightarrow
\pi_{d+1}(G)\rightarrow\pi_{d+1}(G_\infty)\rightarrow\ldots
$$
Hence:
$$\dots\rightarrow\Z\rightarrow\pi_{d+2}(X)\rightarrow\Z_n\rightarrow 0
\rightarrow\ldots
$$
When $\pi_{d+2}(X)$ is known this is enough to give
precise information on the map $\Z\rightarrow\pi_{d+2}(X)$.
In general we only know that  $\pi_{d+2}(X)= \Z^r\oplus T$
where $T$ is torsion and $r$ is a positive integer.

Assuming that the form $\theta_{d+2}=\text{tr}(dg g^{-1})^{d+2}$
vanishes on $G$ (which is
the case if $\pi_{d+2}(G)$ is torsion) we can determine exactly the extension
$\hat{\mathcal G}_e$ through a computation of $\int \theta_{d+2}$ for each
generator in $\pi_{d+2}(X)$ which corresponds to an element in
 $\pi_{1}({\mathcal G}_e).$
An element in the latter group is represented by a map $g:S^1 \times M \to G$
and this map extends to a map $g: D\times M \to G$ which in turn defines
(through canonical projection) a map from $S^2 \times M$ to $X.$ By an
integration
$$\alpha(g) = \left(\frac{-i}{2\pi}\right)^{n+2}\frac{(n+1)!}{(d+2)!}
\int_{D\times M} \text{tr} (g^{-1}dg)^{d+2}$$
we get a real number $\alpha(g)$ and the homotopy class $[g] \in
\pi_1({\mathcal
G}_e)$ is represented by $\exp(2\pi i \alpha(g) )$ in $U(1).$ Thus we are
interested in the values $\alpha(g)$ modulo integers. If all these numbers
are in $\Bbb Z$ then the kernel of the extension $\hat{\mathcal G}_e$ is
represented trivially and there is no global hamiltonian gauge anomaly.
In specific examples we can determine the existence of the global anomaly
without doing any explicit computations. This occurs when $\pi_{d+2}=
\Bbb Z$
and $\pi_{d+1}(G) = \Bbb Z_n.$ In this case we conclude from the exact
homotopy
sequence above that the generator in $\pi_{d+2}(\mathcal G_{\infty})$ is
mapped to $n$ times the generator in $\pi_{d+2}(X)$ and therefore the
value of the integral for the generator in $\pi_{d+2}(X)$ (which is defined
by a generator of $\pi_1 ({\mathcal G}_e)$) is equal to $1/n$ modulo integers.
It follows that $\hat{{\mathcal G}_e}$ is represented faithfully and there
indeed is a global anomaly.

For example the case of $G=SU(3)$ in the fundamental representation
and dim $M =5$ works in this way.
In this case the relevant homotopy is $\pi_6 (G) = \Z_6$ and can be
represented using $\int tr (g^{-1}dg)^7$ on the larger group $SU(4)$
because $\pi_6 (SU(4)) =0$. Using the exact homotopy sequence
$$\pi_7(SU(4))\to \pi_7(SU(4)/SU(3)) \to \pi_6(SU(3)) \to \pi_6(SU(4))$$
gives the exact sequence
$$\Z\to \Z \to \Z_6 \to 0$$
since $SU(4)/SU(3) = S^7$. This shows that the generator of $\pi_6(SU(3))$
gets mapped to 6 times the generator of $\pi_7(S^7).$ Another nice example
is the case of the exceptional simple group $G=G_2$ in the real 7 dimensional
representation. Here one uses the embedding $G_2 \subset SO(7)$ and the fact
that $SO(7)/G_2$ is also equal to $S^7.$ Since $\pi_6(SO(7)) = 0,$
$\pi_6(G_2)
=\Bbb Z_3,$ and $\pi_7(SO(7)) =\pi_7(S^7) = \Bbb Z$ one obtains an
exact
sequence
$$\Bbb Z \to \Bbb Z \to \Bbb Z_3 \to 0$$
of homotopy groups. Thus that the generator of $\pi_7(SO(7))$ is mapped
to three times the generator of $\pi_7(S^7)$ and therefore the (normalized)
integral $\int \TR (dg g^{-1})^7$ corresponding to the elements in
$\pi_6(G_2)$
define the phase factors $1, \exp(2\pi i/3), \exp(4\pi i/3)$ and $\pi_6(G_2)$
is represented faithfully in $U(1).$

\subsection{Algebraic considerations}

In the case of gauge group $G=SU(2)$ and the dimension of the physical
space is three there is a real structure which explains the appearance of
$\Z_2$ determinant bundles (instead of $U(1)$ bundles).

For $2\times 2$ complex matrices there is a real linear (but complex
antilinear) automorphism
$J$ defined by
$$ J \left( \matrix a & b\\ c& d\endmatrix\right)
= \left(\matrix d^* & -c^* \\ -b^* & a^*\endmatrix\right)$$
where star means complex conjugation. Because the spinor field has now
2 internal and 2 space-time components we can think of the Dirac field
as a complex $2\times 2$ matrix function and we can define $J$ as a real
linear operator acting on the Dirac field point-wise in space.
The vector potential acts from the right by matrix multiplication
on $\psi$ whereas the gamma matrices (in this case the Pauli matrices)
act from the left.

The automorphism $J$ has the properties
$J(g) =g \text{ for } g \in SU(2)$ and $J(g) = -g$ if $g$ is hermitean
traceless. From this follows that if $D_A \psi = \lambda\psi$ is an
eigenvector in the external potential $A$ then also $J \psi$ is an eigenvector
corresponding to the same eigenvalue $\lambda.$
For this reason we may choose a real basis of eigenvector $\psi_1,\dots,
\psi_n$ in any given energy range $s < \lambda < t.$ Real means here that
$J\psi_k = \psi_k.$ Any other real basis is obtained by a real orthogonal
transformation $R$ from this basis. Thus the only ambiguity in choosing
a representative in the determinant line is det$R = \pm 1.$ This gives
the required $\Z_2$ structure.
Whether there is an algebraic structure which explains
the global anomaly in other cases remains open.
An interesting test case is the exceptional Lie group $G=G_2$.
The homotopy group $\pi_6$
of $G_2$ is equal to $\Z_3.$
This would lead to $\Z_3$ torsion in the Fock bundle in 5+1 space-time
dimensions for the $G_2$ gauge group (maybe related to quarks in 5+1
dimensions...). There must be some $\Z_3$ structure in the local
determinant bundles over open sets $U_{\l\l'}$ in $\mathcal A$, in the same
way as there is a $\Z_2$ structure in 3+1 dimensions for $SU(2).$

\enddocument